**Bioadhesive Graft-Antenna for Stimulation and Repair of Peripheral Nerves**

Ashour Sliow[1], Zhi Ma[2], Gaetano Gargiulo[3], David Mahns[2], Damia Mawad[4], Paul Breen[3], Marcus Stoodley[5], Jessica Houang[6], Rhiannon Kuchel[7], Giuseppe Tettamanzi[8], Richard D Tilley[7], Samuel J Frost[1], John Morley[2], Leonardo Longo[9] and Antonio Lauto[1,2,3]*.

[1]*School of Science and Health, Western Sydney University, Penrith, NSW, Australia*
[2]*School of Medicine, Western Sydney University, Penrith, NSW, Australia*
[3]*Biomedical Engineering & Neuroscience Research Group, MARCS Institute, Western Sydney University, Penrith, NSW, Australia*
[4]*School of Materials Science and Engineering, University of New South Wales, Kensington, NSW, Australia*
[5]*Faculty of Medicine and Health Sciences, Macquarie University, NSW, Australia*
[6]*Biomedical Engineering, School of Aerospace, Mechanical and Mechatronic Engineering, University of Sydney, Sydney, NSW, Australia*
[7]*Mark Wainwright Analytical Centre, University of New South Wales, NSW, Australia*
[8]*School of Physical Sciences, University of Adelaide, Adelaide, South Australia, Australia*
[9]*Faculty of Human Sciences, University of San Marino, Republic of San Marino*

*Correspondence and requests for materials should be addressed to A.L. (email: a.lauto@westernsydney.edu.au)*

**Abstract**

Peripheral nerve injuries are difficult to treat due to limited axon regeneration; brief electrical stimulation of injured nerves is an emerging therapy that can relieve pain and enhance regeneration. We report an original wireless stimulator based on a metal loop (diameter ~1 mm) that is powered by a transcranial magnetic stimulator (TMS). The loop can be integrated in a chitosan scaffold that functions as a graft when applied onto transected nerves (graft-antenna). The graft-antenna was bonded to rat sciatic nerves by a laser without sutures; it did not migrate after implantation and was able to trigger steady compound muscle action potentials for 12 weeks (CMAP ~1.3 mV). Eight weeks post-operatively, axon regeneration was facilitated in transected nerves that were repaired with the graft-antenna and stimulated by the TMS for 1 hour/week. The graft-antenna is an innovative and minimally-invasive device that functions concurrently as a wireless stimulator and adhesive scaffold for nerve repair.

Injured peripheral nerves have the capacity of regenerating, however, functional recovery fails unless transected nerves are surgically repaired to allow axon regrowth into the distal nerve stump [1,2] where Schwann cells support their growth [3]. The outcome of surgical anastomosis has limitations too, for example, when axon growth must cover a large gap to restore nerve continuity and functionality [4]. A common procedure for reconnecting transected nerves is bridging the large gap with a nerve graft harvested from the patient (autograft) or using non-nervous grafts such as biological and synthetic conduits [5]. Grafts are typically connected to tissue by sutures that are invasive and may cause detrimental side effects such as scarring and inflammation [6,7]. The limitations of current surgical techniques have prompted the search for alternative therapies. Very encouraging results have been recently reported in studies where injured peripheral nerves were electrically stimulated which led to pain relief, better nerve regeneration and functional recovery than untreated nerves [8,9]. The electric pulses typically have an amplitude of a few volts, duration of few hundred microseconds and low frequency (1-50 Hz, number of pulses per second) [10]. A stimulation time of 1 hour after nerve repair has been reported to have a similar regenerating effect to longer stimulation regimes [11]. The molecular mechanism of axon regeneration following brief electrical stimulation is currently under investigation; studies have nonetheless highlighted that the stimulation elevates the level of neuronal cyclic adenosine monophosphate (cAMP) [12–14], accelerates and increases the upregulation of neurotrophic factors and their receptors in neurons [8,11] and Schwann cells [15–17]. Weiner et al published a study where a miniaturized wireless neurostimulator was transforaminally placed at the dorsal root ganglion of patients to treat low back pain [18]. The system included an implantable stimulator with 4 electrodes and an externally worn transmitter to power the device that was positioned ~ 6 cm from the receiver. Electrical stimulation caused an overall pain reduction of ~60% in patients. A similar device was also implanted adjacent to facial nerves in patients suffering from chronic pain [19]. Electrical stimulation with a 16-electrode array system proved to modulate the spinal circuitry and enabled full weight-bearing standing in a patient who suffered spinal cord injury [20]. Hentel tested a wireless stimulator that efficiently delivered constant-voltage pulses to rat sciatic nerves from a step-up voltage regulator under microprocessor control [21]. MacEwan and co-workers fabricated flexible wireless stimulators that were implanted subcutaneously into rats. Thin-film implants delivered brief electrical stimulation to sciatic nerves after nerve crush or nerve transection-and-repair injuries. The treatment significantly increased the rate of functional recovery in both animal groups [22]. The stimulators consisted of 3 components: a receiver coil with a demodulating circuit, microwire leads and a silicone nerve cuff. Other implantable wireless devices have been proposed to simultaneously record neural activities and stimulate nerves while sharing a single cuff electrode [23]. Despite some operative success, the devices so far described are still bulky and have various degrees of complexity in circuit design. They are usually encapsulated by an inert material to prevent tissue toxicity and are implanted using invasive means, such as sutures, to avoid migration from the operation site. Stimulators are typically not designed to function concurrently as scaffolds and grafts for injured nerves as their role is limited to supplying electrical pulses to neural tissue [24]. In this report, we describe an innovative wireless stimulator based on a metal loop, and powered by a transcranial magnetic stimulator. The loop is integrated in a chitosan scaffold that functions as a graft when applied to transected nerves. This scaffold is anchored to tissue by a laser without sutures, thus exploiting the photo-adhesive properties of the scaffold, which was previously developed by our group and tested on intestine, cardiac tissue and peripheral nerves [25–29]. We have named this device "graft-antenna" to highlight the double functionality of the implant.

# Results

*Characterization of the TMS-antenna system*

The loop antenna successfully coupled to the transcranial magnetic stimulator (Figure 1) that induced voltages of the order of $10^{-2}$ - $10^{-3}$ volts in the antenna. In particular, the voltage induced in the loop antenna decreased as the TMS coil was moved away along the Z-axis (Figures 2a, 2b); when stimulation occurred at ~0.72 T (60% $B_{max}$), the voltage varied from 18 to 2 mV as the distance increased from 10 to 150 mm respectively. The voltage induced in the antenna was symmetric around the origin, indicating that the B-field received by the antenna was symmetric with respect to the axes origin, as described by the TMS manufacturer. In this instance, the voltage values decreased as the antenna moved away from the centre of the TMS coil along the X-axis and Y-axis (Figures 2c, 2d).

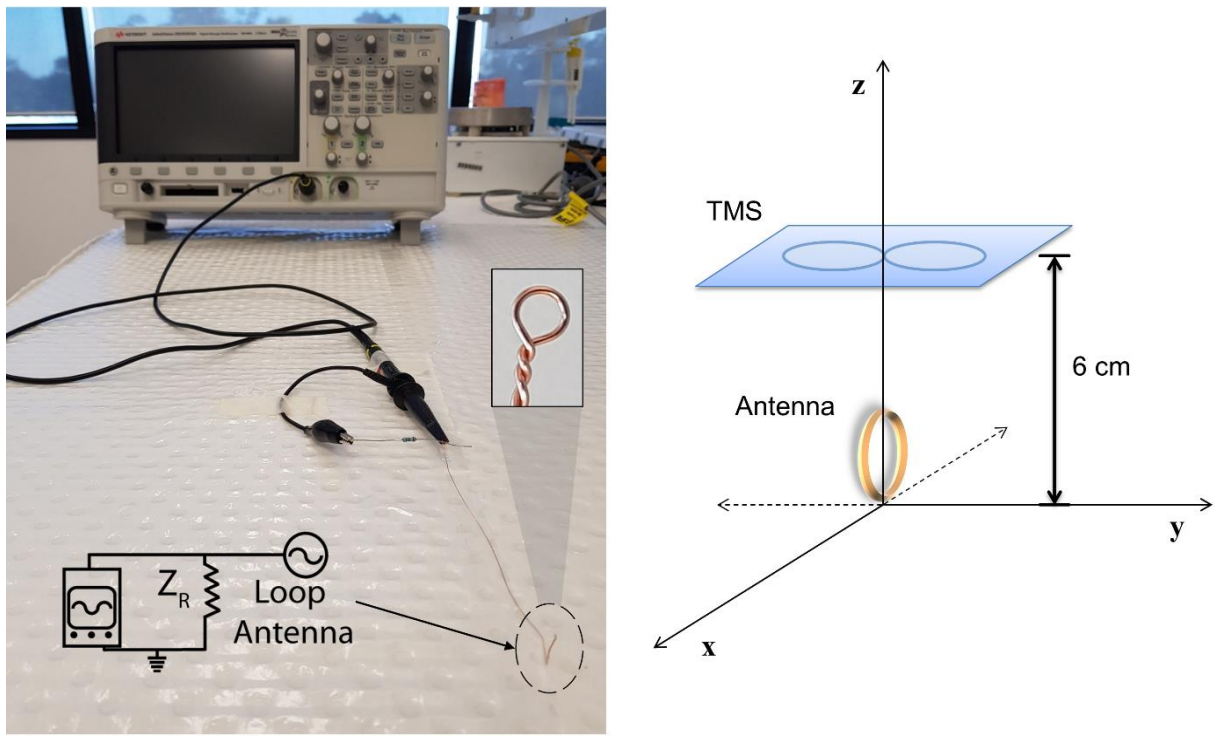

**Fig. 1** (Left) Experimental set-up of the voltage measurements in the copper loop antenna using the oscilloscope and TMS. (Right) Illustration of the TMS and loop antenna positions in Cartesian coordinates; their relative position varied along the three axes during the antenna characterization experiments. The optimal TMS position for nerve stimulation in rats was found to be above the animal at (0,0,6 cm) when the antenna was placed around the nerve in the XY plane (at the axes origin).

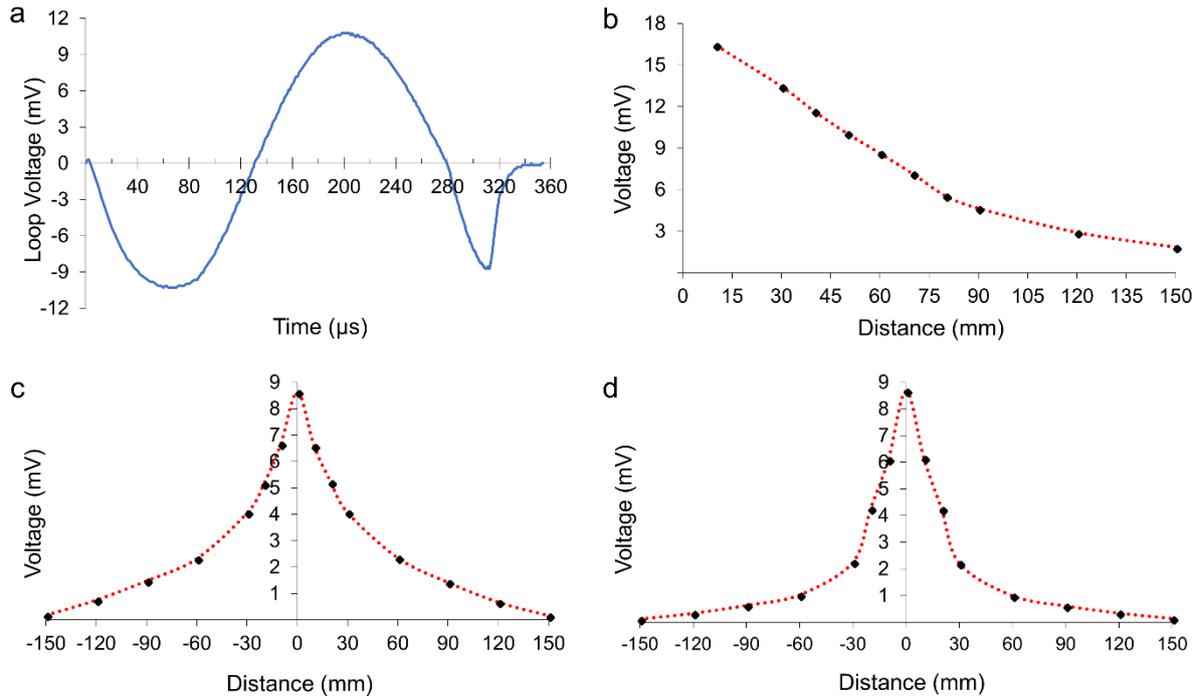

**Fig. 2** (a) Plot of the voltage vs time that is induced in the copper loop antenna by the TMS. The duration of the voltage pulse is ~ 350 µs. (b) Voltage induced in the loop antenna vs distance when the TMS is positioned on the z-axis and the antenna is fixed at the axes origin; (c) Voltage vs distance when the antenna moves along the x-axis (x,0,0) and the TMS height is fixed at (0,0,6 cm). (d) Voltage vs distance when the antenna moves along the y-axis (0,y,0) and the TMS height is fixed at (0,0,6 cm). The symmetry of the voltage with respect to the x- and y-axes is due to the symmetric magnetic field generated by the TMS coil. Three independent experiments were performed for each plot and each point value is the average of ten measures. The standard deviation (≤ 0.06 mV) is not shown in the plots for image clarity.

*Loop-antenna stimulation of nerves*

After the characterization of the TMS-antenna system, we tested if this system could stimulate tissue. The TMS triggered compound action potentials either in muscles (0.67 ± 0.09 mV, n = 3, Supplementary Table 1) or nerves (0.33 ± 0.05 mV, n = 3) when the loop antenna was placed around the sciatic nerve of rats; no action potential was elicited during TMS stimulation without the loop antenna (Figure 3). It is of note that eliciting an action potential without the antenna was impossible, even when the TMS coil was positioned just above the nerve (z = 3 cm, y = 0, x = 0) at maximum magnetic field amplitude ($B_{max}$ ~1.2 T). The action potentials were triggered only in the presence of the loop antenna when the B-field was ≥ 0.6 T (50% $B_{max}$), while no muscle or nerve response was recorded at lower magnetic magnitudes. If the B-field was ≥ 0.84 T (70% $B_{max}$), the movement of the rodent's body, triggered at each pulse, affected the electrode stability and reliability of measurements; it was also observed that when the magnetic field was ≥ 0.94 T (80% $B_{max}$), the action potential was obscured by an artefact due to the intensity of the TMS pulse. A distinctive twitch of the leg was nonetheless observed in all these cases (Table 1). The compound action potentials in muscles and nerves did not change significantly when the antenna around the nerve was earthed to the oscilloscope. No action potential was elicited at any B-field magnitude, including $B_{max}$, when the antenna wrapped by a plastic coating (n = 3); the latter experimental outcome suggests that contact between the copper loop and nerve was necessary for stimulating nerves. It was also noted that during TMS irradiation, action potentials were occasionally not triggered; this "misfiring" was due to a small displacement (~10 µm) of the loop antenna around the nerve, caused by the leg twitch. The least misfire occurred at 0.72-0.84 T (60-70% $B_{max}$) where no action potential was elicited at 8% of the time, while at 80% $B_{max}$, artefacts obscured any possibility of an action potential recording. These results indicate that the TMS coil should be positioned ~6 cm from the rat nerve, and function at 60% $B_{max}$ in order to ensure stable and reliable measurements of action potentials (Figure 1). At these settings, the current induced by the TMS in the loop antenna positioned around the nerve had a peak value of 3.2 ± 0.3 µA; a value of 3.6 ± 0.1 µA was measured in the antenna without the nerve (p = 0.0714, unpaired t-test, n = 3).

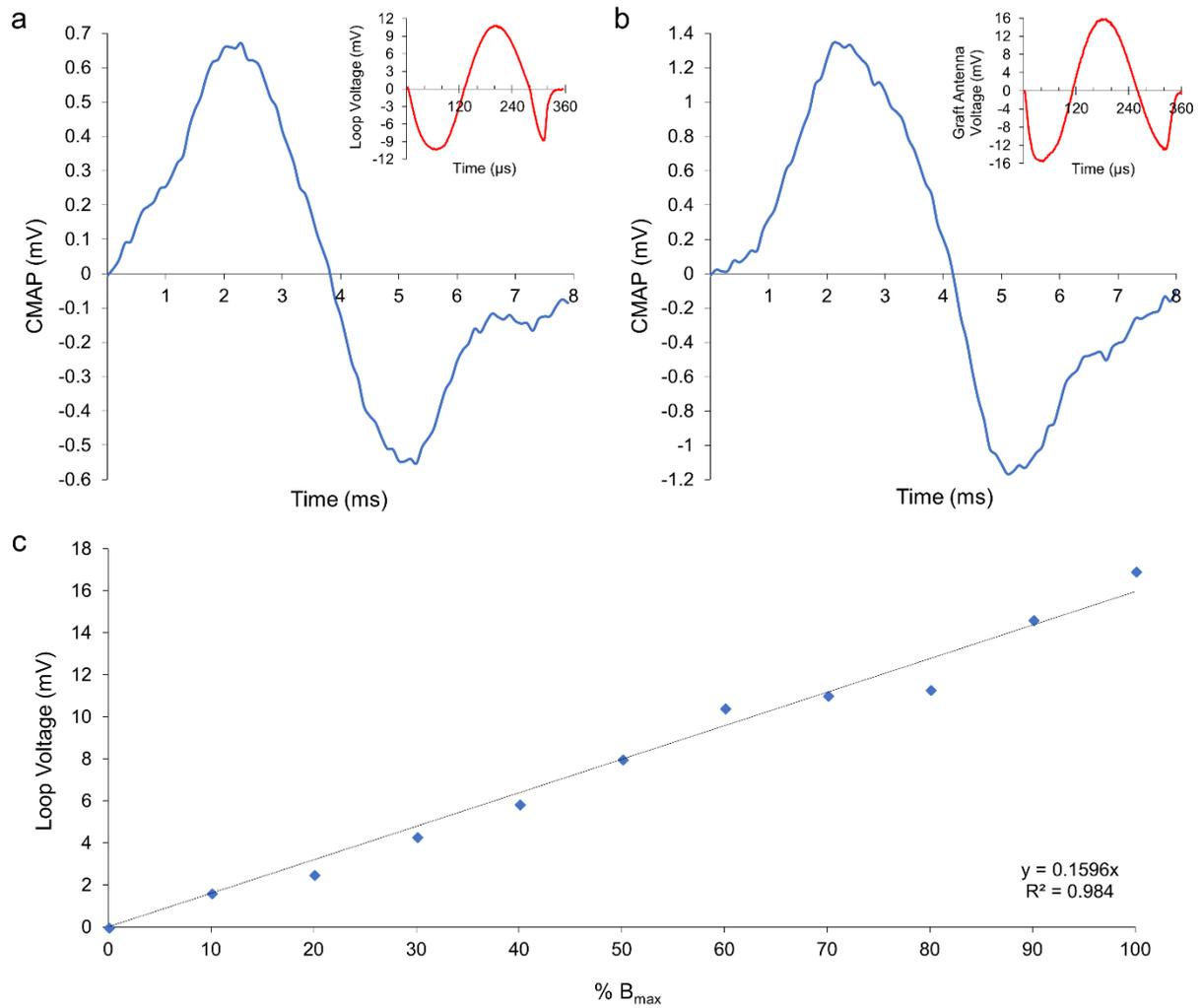

**Fig. 3** (a) A typical CMAP that is triggered by the copper loop antenna placed around the sciatic nerve (n = 3). (b) A typical CMAP triggered by the gold loop embedded in the graft-antenna when it is laser-bonded around the sciatic nerve (n = 3). The TMS was located 6 cm above the loop antenna and nerve (x = 0, y = 0, z = 6 cm), and irradiated by an electromagnetic field with intensity B ~0.72 T (60% $B_{max}$). The voltage pulse induced in the loop antenna by the TMS is shown in the figure insets. (c) The voltage amplitude of the copper loop is proportional to the magnetic field magnitude, showing the inductive nature of this antenna.

**Table 1. Compound Muscle Action Potential vs Magnetic Field Amplitude**

| % $B_{Max}$ (Magnitude) | CMAP (mV) | Loop Voltage (mV) | Misfire (%) | Muscle twitch |
|---|---|---|---|---|
| 10% | - | 1.63 ± 0.04 | - | No |
| 20% | - | 2.52 ± 0.03 | - | No |
| 30% | - | 4.32 ± 0.02 | - | No |
| 40% | - | 5.87 ± 0.03 | - | No |
| 50% | 0.23 ± 0.06 | 8.01 ± 0.05 | 12% | Yes |
| 60% | 0.72 ± 0.07 | 10.43 ± 0.07 | 8% | Yes |
| 70% | 0.86 ± 0.04 | 11.03 ± 0.05 | 8% | Yes |
| 80% | - | 11.31 ± 0.09 | - | Yes |
| 90% | - | 14.64 ± 0.03 | - | Yes |
| 100% (1.2 T) | - | 16.94 ± 0.05 | - | Yes |

Legend. *% $B_{Max}$*, amplitude % of maximum TMS magnetic field; *CMAP*, compound muscle action potential triggered by the copper loop antenna that was irradiated by the TMS; *Loop Voltage*, voltage induced in the copper loop antenna by the TMS; *Misfire*, failure percentage in triggering CMAPs following TMS irradiation. It was noted that when contact was (partially) lost between the loop antenna and the nerve, no action potential was elicited. *Muscle Twitch*, visual recognition of muscle twitches during TMS stimulation. Three independent experiments were performed and for each experiment the CMAP or voltage values were an average of 30 measures.

*Compound Muscle Action Potential: AC versus DC currents*

This test gauged the amount of DC current required to trigger a CMAP similar to that elicited by TMS-antenna stimulation. It was found that when a DC current of ~18µA was injected in the nerve by a standard stimulator, a CMAP with an amplitude of 0.88 ± 0.10 mV of was triggered, while the TMS elicited a CMAP of 0.85 ± 0.10 mV in the same nerve, generating an AC peak current of 3.7 ± 0.1 µA inside the loop antenna. The CMAP amplitudes originated by the two methods were not statistically different (p = 0.5142, paired t-test, n = 4). Notably, the CMAP decreased sharply to 0.53 ± 0.04 mV when a DC current of ~15 µA was injected in the nerve, showing a clear trend between current and amplitude (Supplementary Figure 1). In the previous experiments, copper loops were implanted in live rats for ~15 minutes and were used as a proof of concept considering the moderate cytocompatibility of copper [30]. Note that loop antennas made of biocompatible materials such as titanium or gold are indicated for long term implantation and stimulation of nerves without any toxic side effects.

*In vivo implantation of graft-antennas: stimulation of uncut nerves*

A biocompatible device that functions as a sutureless scaffold and wireless stimulator at the same time is sought to repair nerves and facilitate axon regeneration. With this purpose in mind, we designed and developed a graft-antenna that comprised of two parts, namely a chitosan-based film with photo-adhesive capability and a thin loop of gold plated on the adhesive (Figure 4). The graft-antenna was photochemically bonded by a laser to the sciatic nerve and the TMS elicited CMAPs for a 12-week period (Figure 5a); the device did not migrate during this time. The graft-antenna triggered steady action potentials during the implantation time and remarkably, the CMAPs (~1.3 mV) recorded before euthanizing animals were not statistically different from the values measured immediately after implantation ($p > 0.05$, one-way ANOVA, Tukey's multiple comparison test, n = 5). This result demonstrates that the graft-antenna is stable once implanted and is a reliable stimulator that does not cause detrimental effects on nerve conduction at the selected stimulation regime. The action potentials recorded when the leg wound was closed (weeks 2 -11) varied between 0.8 and 1.1 mV and were lower than the ones recorded on the exposed nerves (immediately after implantation and before euthanasia – week 1 and week 12); this can be ascribed to the skin and muscles covering the nerves and attenuating the electromagnetic field over the gold loop of the graft-antenna ($p < 0.05$, one-way ANOVA). Histology results showed that there was no statistical difference in axon number, fiber diameter, axon diameter, and myelin thickness between the contralateral (control), proximal and distal sites of the nerve stimulated with the graft-antenna (Figure 5b, Supplementary Table 2). This outcome agrees with the electrophysiology data and confirms the graft-antenna is as a safe stimulator. Of note is the fact that the graft-antenna elicits significantly higher CMAPs than copper loops when stimulated with the same magnetic magnitude ($1.43 \pm 0.11$ mV and $0.69 \pm 0.09$ mV respectively; $p = 6.72123E-05$, unpaired t-test, n = 5). The currents measured in graft-antennas were also significantly higher than those in copper loops ($13.5 \pm 0.2$ µA vs $3.2 \pm 0.3$ µA; p = 8.98729E-08, unpaired t-test, n = 3). This may be due to the gold strip altering the geometry and enlarging the diameter of the loop when the adhesive is wrapped around the nerve and a section of it is superimposed (~25%). Remarkably, no misfire occurred during the nerve stimulation with the graft-antenna, indicating that this device was more stable around the nerve than the copper loop.

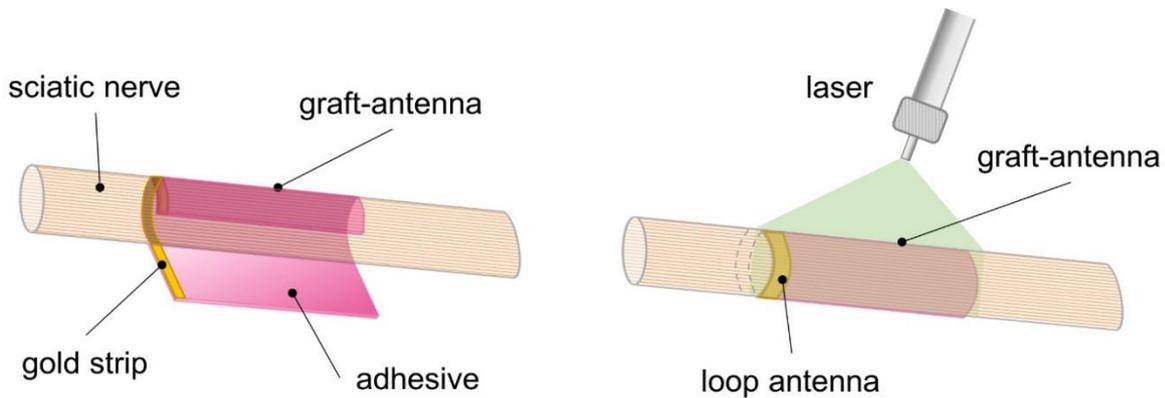

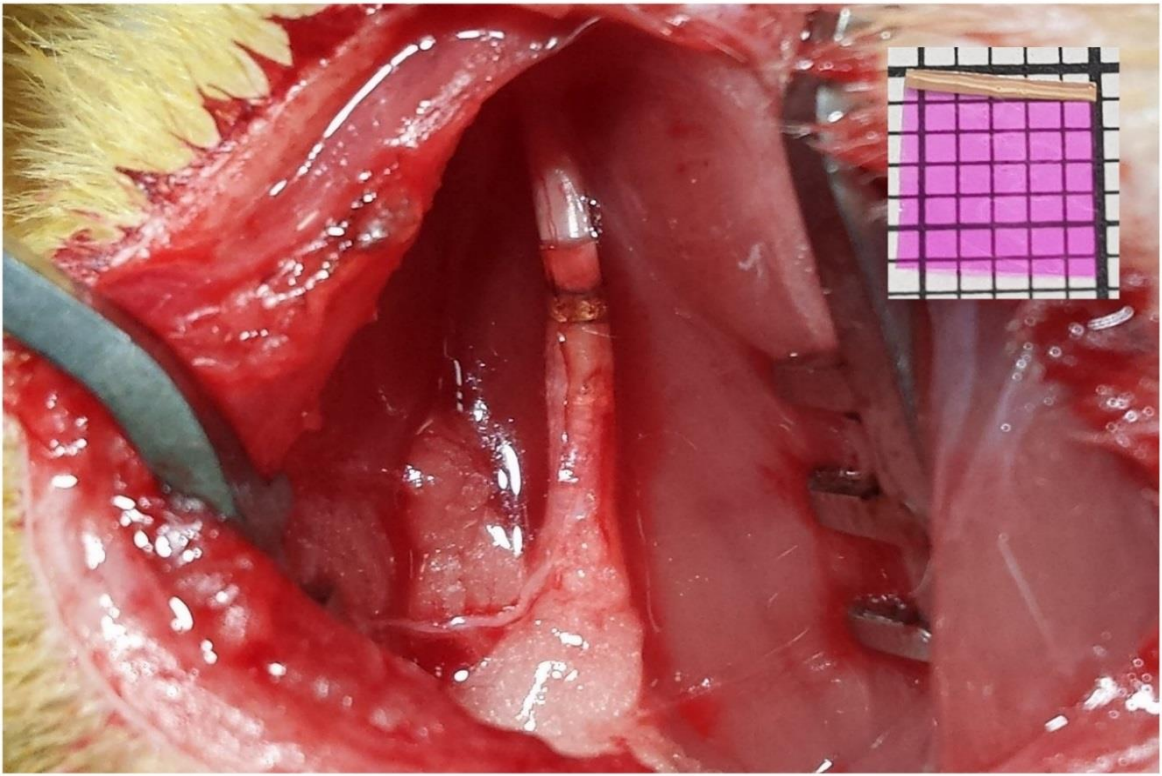

**Fig. 4** *In vivo* image of the graft-antenna after being laser-bonded to the intact sciatic nerve (diameter ~1 mm) of a rat. The graft-antenna is made of a chitosan-based film containing the dye rose bengal and a strip of gold (thickness ~70 nm) embedded in the adhesive (inset). When the film is placed around the nerve, the gold strip becomes a loop antenna. The green laser ($\lambda$ = 532 nm) irradiates and activates the rose bengal inside the adhesive film that bonds to tissue. Note that the blood vessels do not appear coagulated or otherwise damaged under or in the proximity of the graft-antenna. The gold loop antenna contacts and surrounds the nerve underneath the graft that conforms to the nerve without constriction.

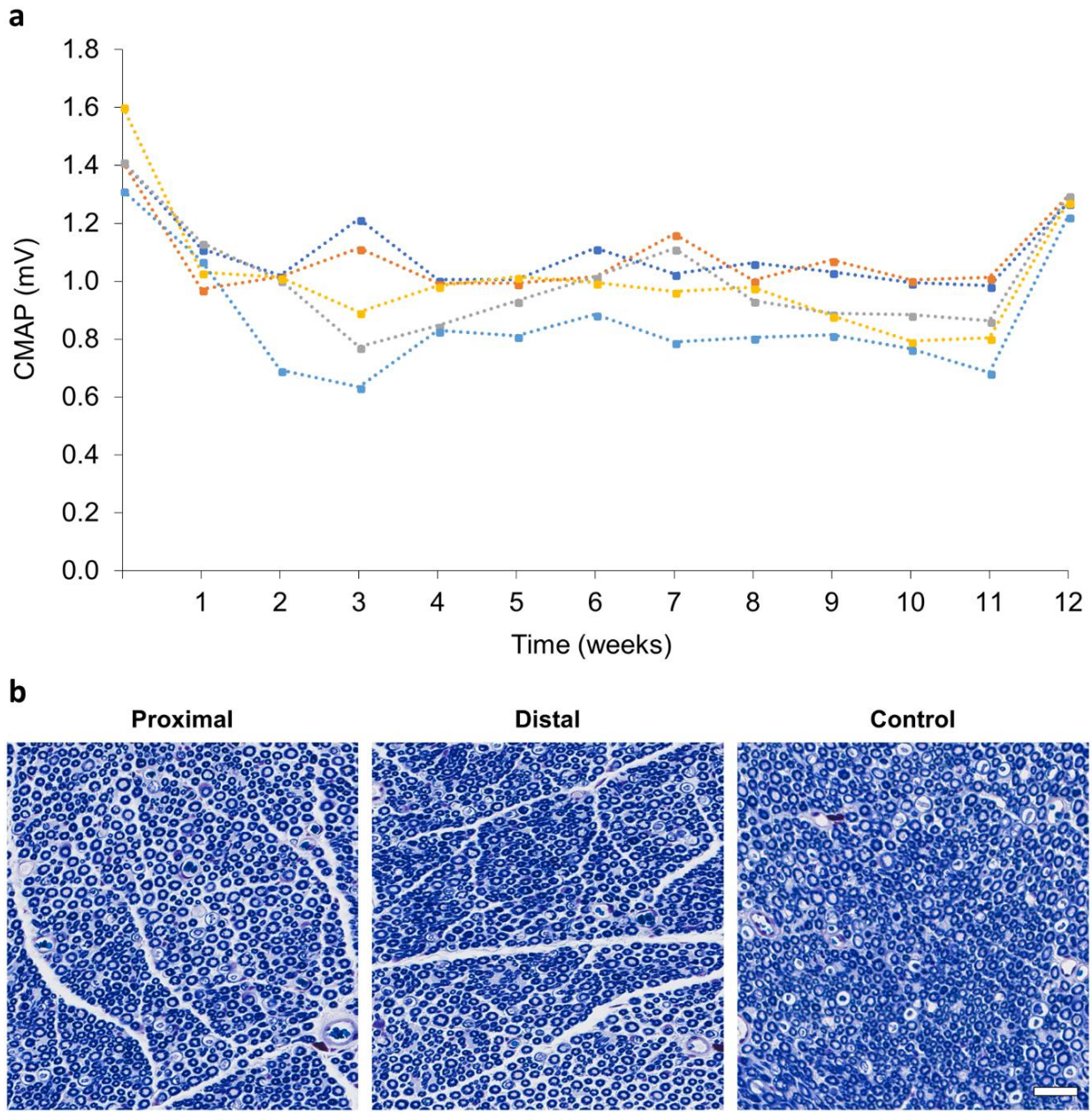

**Fig. 5** (a) Plot of the CMAP amplitude that is triggered in healthy (uncut) sciatic nerves by the graft antenna during TMS stimulation, over a 12-week period. The graft-antenna was firstly bonded to the nerve by a green laser, followed by TMS irradiation of the graft-antenna once a week for 1 hour during a 12-week period. The CMAPs at week 0 and week 12 were not statistically different indicating that the graft-antenna is a reliable and stable stimulator ($p > 0.05$, one-way ANOVA, Tukey's multiple comparison test). The CMAP between weeks 1 and 11 was fairly constant, although lower than the CMAP at week 0 and 12 ($p < 0.05$, one-way ANOVA, Tukey's multiple comparison test). This was due to the electromagnetic shielding of tissue over the nerve: the CMAP was recorded with the nerve exposed to the TMS only at week 0 and 12. The standard deviation ($\leq 0.2$ mV) is not shown in the plot for visual clarity. Each point represents the average of 10 pulses. (b) Histologic slides from proximal and distal sites of the nerves to which the graft-antenna was laser bonded. These nerves were briefly stimulated for 1 hour by the TMS via the graft-antenna over a 12-week period (n = 5). The TMS stimulation and laser bonding procedure did not alter or affect axon number, diameter and myelin thickness when compared to untreated contralateral nerves ($p > 0.05$, one-way ANOVA). (Scale bar = 10 µm).

*In vivo nerve grafting of transected nerves*

The aim of this experiment was to test the graft-antenna *in vivo* and assess its ability of stimulating axon regrowth without adverse effects after nerve transection and grafting. All grafted nerves were patent and in continuity at 8 weeks post-operatively; no sign of uncharacteristic inflammation or tissue growth was macroscopically evident at the operation site (Figure 6). The nerves that were repaired and TMS-stimulated with the graft-antenna produced a muscle twitch at weeks 6, 7 and 8, signalling regeneration of axons through the graft. Before euthanasia at week 8, the CMAP of animals implanted with the graft-antenna was measured with a value of ~48% of the un-operated contralateral nerves ($0.33 \pm 0.09$ vs $0.68 \pm 0.09$ mV, n = 5, Figure 7); while animals grafted with the adhesive alone elicited CMAPs that were $0.26 \pm 0.09$ mV or ~38% of the contralateral nerves (Table 2). A similar trend was observed for the compound nerve action potentials (CNAPs) of the graft-antenna and adhesive-only groups that were 71% and 41%, respectively, of the contralateral nerves ($0.32 \pm 0.01$ mV). The nerve conduction velocity (NCV) of nerves repaired with the graft-antenna and adhesive were 71% and 61%, respectively, of the contralateral nerves ($58.7 \pm 3.1$ m/s). When compared to the adhesive-only repairs, the graft-antenna group had higher CMAPs ($p < 0.001$), CNAPs ($p < 0.001$) and NCVs ($p < 0.01$) (one-way ANOVA, Tukey post-test). The histology analysis (Figure 7) showed that all distal nerve sections were populated with axons; axon count of distal sites ($1396 \pm 68$) operated with the graft-antenna was significantly higher than distal sites operated with the adhesive only ($1202 \pm 66$) ($p = 0.0363$, one-way ANOVA, Tukey post-test). However, myelin thickness, fiber and axon diameter were not significantly different in the two groups (Supplementary Table 3). Contralateral nerves had the highest number of myelinated axons as expected ($1861 \pm 79$, $p < 0.0001$, one-way ANOVA).

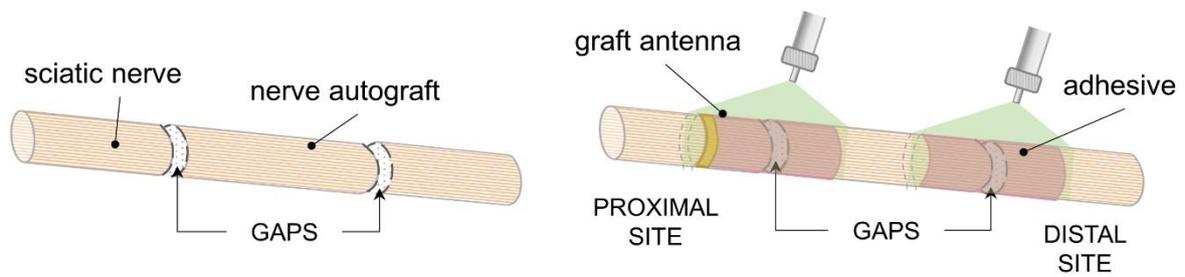

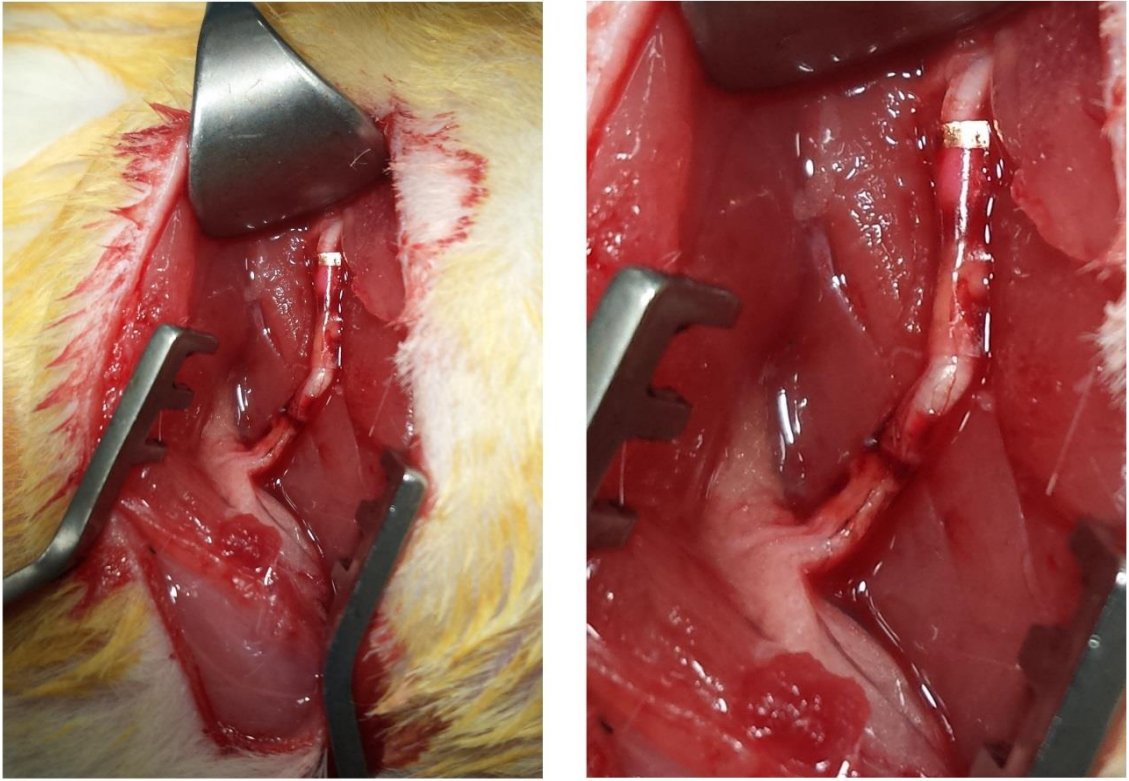

**Fig. 6** (Left) *In vivo* image of the sciatic nerve after laser-grafting. The nerve graft is reconnected to the distal stump by the adhesive and to the proximal stump by the graft-antenna; the green laser photo-crosslinked the adhesive to tissue. (Right) The gaps between the nerve graft and the stumps are noticeable in the magnified image. Nerves and surrounding tissue do not appear damaged by the laser procedure.

**Table 2. Electrophysiological Parameters of Grafted Nerves**

|  | **CMAP (mV) TMS Stimulation** | **CMAP (mV) DC Stimulation** | **CNAP (mV) DC Stimulation** | **NCV (m/s)** |
|---|---|---|---|---|
| **Graft – Antenna** | 0.98 ± 0.09 | 0.33 ± 0.09 | 0.21 ± 0.08 | 41.74 ± 4.13 |
| **Chitosan Adhesive** | - | 0.26 ± 0.09 | 0.14 ± 0.07 | 36.65 ± 6.33 |
| **Control** | 1.43 ± 0.11 | 0.69 ± 0.09 | 0.32 ± 0.08 | 59.28 ± 6.51 |

The electrophysiological parameters for sciatic nerves 8 weeks post-operatively. Nerves repaired with the graft antenna (n = 5) had higher values of compound muscle action potential (CMAP), compound nerve action potential (CNAP) and nerve conduction velocity (NCV) than nerves repaired with the chitosan adhesive only (one-way ANOVA, Tukey post-test, $p < 0.01$). Legend: *DC Stimulation*, nerves stimulated with a DC stimulator; *TMS Stimulation*, nerve stimulated with the TMS.

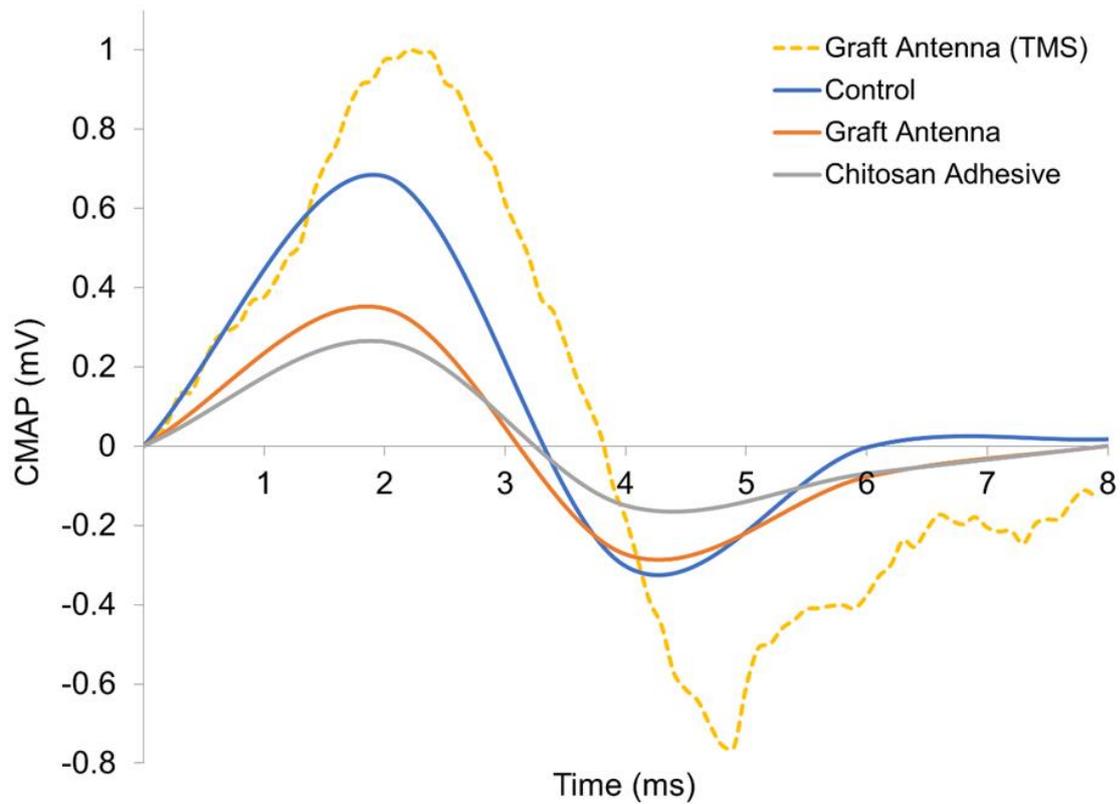

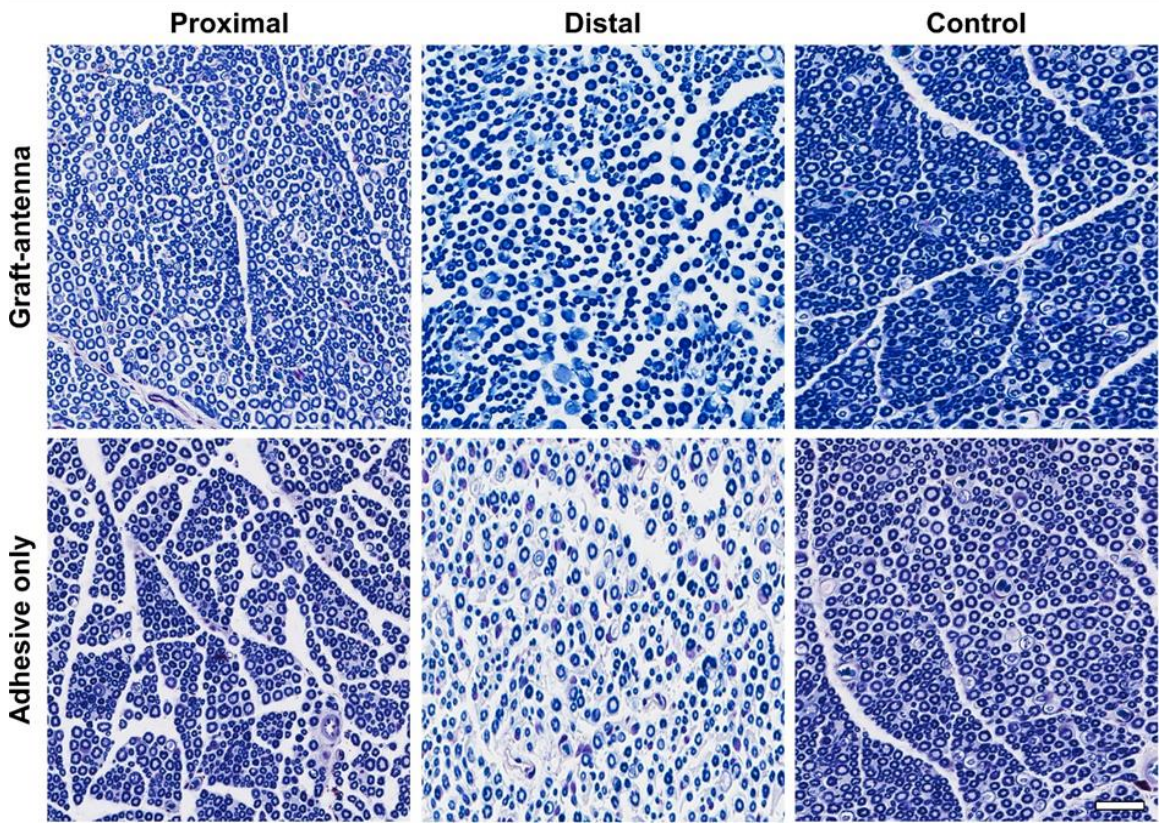

**Fig. 7** (Top) Typical CMAPs of nerves 8 weeks post-operatively; these nerves were repaired either with the graft-antenna or adhesive only and underwent TMS stimulation once a week for 1 hour. The amplitude of the CMAP was larger in nerves repaired with the graft-antenna (p < 0.01, one-way ANOVA, Tukey's multiple comparison test, n = 5). All action potentials were elicited by a DC

stimulator, except for the one represented by a dot line that was triggered by the TMS. (Bottom) Histologic slides from proximal and distal sites of the grafted nerves that were repaired with the graft-antenna (first row) and adhesive only (second row). The unoperated contralateral nerves served as control in both groups. Subsequent to the operation, the nerves were briefly stimulated for 1 hour by the TMS over a period of 8 weeks (n = 5). After 8 weeks, a larger number of myelinated axons regrew through the distal site of the nerve graft repaired with the graft-antenna than the site repaired with the adhesive only (1396 ± 68 vs 1202 ± 66, $p = 0.0363$, one-way ANOVA, Tukey's multiple comparison test). Axon diameter and myelin thickness were not statistically different among the two operated groups ($p > 0.05$, one-way ANOVA). The histomorphometric parameters were significantly higher in the control group (myelinated axons = 1861 ± 79, $p < 0.0001$, one-way ANOVA). (Scale bar = 10 µm).

**Discussion**

Repairing and electrically stimulating peripheral nerves with a non-invasive device is very challenging and the current scientific and technological know-how has yet to produce an effective system to combine and perform these two tasks together. We succeeded in fabricating and testing *in vivo* a biocompatible graft-antenna that repairs peripheral nerves without sutures and stimulates them wirelessly without inserting electrodes in the body. The graft-antenna was implanted around the uncut sciatic nerves of rats for 12 weeks and showed remarkable consistency in triggering CMAPs for the entire duration of the experiment. The histology results also indicated that the stimulation regime adopted in this study (1 hour per week, pulse duration ~350 μs, repetition rate = 1 s, $B_{max}$ ~0.72 T) did not alter the number and morphometric parameters of axons when compared to untreated contralateral nerves. The brief TMS stimulation of transected nerves that were repaired with the graft-antenna facilitated axon regeneration and histological analysis revealed that the axon count of the distal site was 76% that of unoperated contralateral nerves. Nerves grafted with the adhesive only were also stimulated by the TMS once a week but their axon count and electrophysiology performance were inferior.

The principles underlying the TMS-antenna stimulation of tissue can be inferred. The voltage generated in the loop antenna by the TMS sets up currents, namely ~3.6 μA in the copper loop and ~13 μA in the gold loop of the graft-antenna. These currents create propagating electromagnetic fields both at the surface of and outside the loop, to which the electric field is purely tangential [31]. When the loop antenna surrounds the sciatic nerve, there is contact at the interface between the loop antenna and the nerve; the tangential electrical field is (in first approximation) identical at the interface due to Maxwell's boundary conditions. This tangential field around the nerve perimeter produces the voltage responsible for triggering the action potentials (Figure 8). In fact, our experiments demonstrated that no action potential is elicited if the loop antenna is shielded with a plastic insulator during TMS irradiation. Therefore, the contact between the loop antenna and nerve tissue is necessary to elicit action potentials. It can be ruled out that action potentials are provoked by electromagnetic fields traveling inside the nerve that have been generated by the loop currents. These secondary fields can indeed travel into nerves through the plastic shield but they have no stimulatory effect. Note that naked copper loop antennas (no plastic shield) also failed in triggering action potentials whenever the contact with the nerve was lost, even if the loop remained very close to nervous tissue (~10 µm) as recorded during the loop antenna misfires. This confirms the hypothesis that the secondary electromagnetic fields inside the nerve do not trigger action potentials. It appears unlikely that action potentials are elicited by currents flowing into the nerve from the loop antenna. Our results show that CMAPs with an amplitude of ~0.85 mV are triggered when currents in the copper loop antenna are $3.2 \pm 0.3$ µA; if the nerve is removed from the loop, the antenna current increases to $3.6 \pm 0.1$ µA. If we assume that a 0.4 µA current flows into the nerve from the antenna, no action potential can be elicited because a DC current of 18 µA is necessary to initiate a 0.85 mV action potential. Furthermore, the amplitude of action potentials drops from ~0.85 to ~0.53 mV when the DC injector delivers a 15 µA current into the nerve, which is still much larger than 0.4 µA (Supplementary Figure 1). Although details of axon depolarization are still unclear, our results indicate that the voltage around the loop-nerve interface elicits action potentials without the intervention of secondary electromagnetic fields or leaked currents from the metallic loop into the nerve. A similar result is obtained using the graft-antenna as the currents measured when the nerve is or isn't inserted in the graft are ~13.5 and ~15.2 µA, respectively. The CMAPs elicited by the graft-antenna were larger than the ones triggered by the copper loop antenna in the uncut sciatic nerves (~1.4 vs ~0.7 mV); this can be ascribed to the higher voltage produced at the nerve interface by the graft-antenna. This hypothesis is in agreement with the observation that currents in the graft-antenna were higher than the ones induced in the copper loop. Variable currents indeed contribute in generating tangential electric fields at the nerve interface; Li *et al* reported that the tangential near fields of a small loop antenna are proportional to the loop current [31].

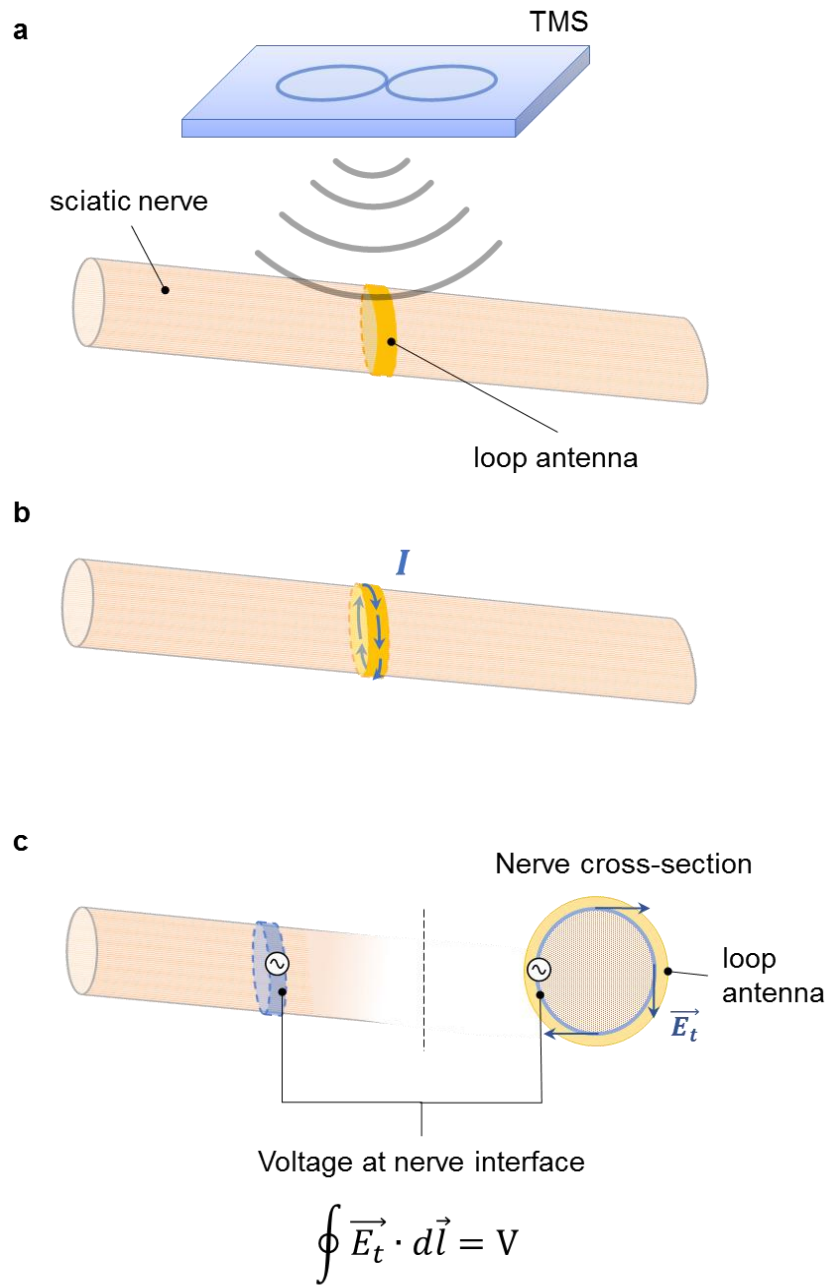

**Fig. 8** Schematic of CMAP activation in the sciatic nerve. (a) The TMS irradiates the nerve and sets up currents in the loop antenna (b); these currents generate tangential electrical fields at the nerve interface. (c) The electrical field circulation around the nerve circumference is the voltage that triggers the compound muscle action potential. The loop antenna acts as an electrical stimulator that applies voltage all around the nerve upon TMS activation.

The regime of stimulation selected for this study was based on previous relevant reports [20,32,33]. Al-Majed *et al* [11] demonstrated that a 1-hour stimulation accelerated axonal regeneration (pulse duration ~100 µs, 3 V, 20 Hz) and Brushart *et al* [34] found that a similar stimulation can enhance regeneration specificity of sensory axons. In a clinical study, a 16-electrode array was implanted over the spinal cord dura. An average stimulation period of 54 minutes was applied with biphasic rectangular pulses ranging from 210 to 450 µs at 0.5-10 V and frequency of 5-40 Hz. Results proved that electrical stimulation could modulate the physiological state of the spinal circuitry in a patient who suffered clinically motor complete spinal cord injury [20]. Another report suggested that bipolar stimulation produces larger spinal motor-evoked potentials and better motor function outcome than monopolar stimulation [35]. The beneficial effect of brief electrical stimulation was also shown in patients who underwent carpal tunnel release surgery [36]. The stimulators reported above are implanted with conventional means; electrodes, for example, are routinely sutured to tissue. Sutures are notoriously invasive and the incidence of complications is high (30-40%). The commonest hardware-related complications are electrode migration, electrode failure and fracture, while other frequent side-effects include infection and pain over the implant [37]. The graft-antenna is a biocompatible device that is fixed to tissue without sutures and with non-invasive means, namely by a low power laser that keeps the tissue temperature below 37 ºC during irradiation [38,39]. Our study has shown that the graft-antenna is stable over the nerve and does not shift position or diminish in its stimulation capacity over a 3-months period (Figure 5). The graft-antenna is also powered by a TMS at levels of radiation that are safe to patients [40]. Furthermore, the components of the device; namely chitosan and a small strip of gold, are non-toxic and biologically compatible, with the chitosan film also biodegradable [41]. Future studies are foreseen to compare axon regeneration when nerves are repaired using conventional sutures or with the graft-antenna device. In conclusion, an innovative antenna-TMS system has been developed and characterized for nerve stimulation. The antenna can be a biocompatible metal loop such as gold or titanium positioned around the nerve, or a graft-antenna based on chitosan and comprising a gold loop. The graft-antenna is at the same time a wireless stimulator powered by a TMS, and an adhesive scaffold that is fixed to tissue by light and without sutures. The graft-antenna is stable in the body after implantation and can facilitate axon regeneration with no significant adverse effects.

**Methods**

*Characterization of the TMS-antenna system*

A TMS irradiated a small loop antenna inducing a voltage that was measured with varying relative positions of the TMS and loop. Copper loop antennas were fabricated with a diameter of 1.0 ± 0.1 mm and wire thickness of 0.10 ± 0.01 mm. The antenna was connected to a resistor (1 kΩ) and a coaxial cable (Z = 50 Ω) that was plugged to the earth terminal of an oscilloscope (DSOX2014A, 100MHz and 500 Gsample/s), as illustrated in Figure 1. The antenna and the coaxial cable were 1 meter apart to prevent electromagnetic coupling. When the electromagnetic pulse irradiated the loop antenna, the variable magnetic field induced a voltage that was measured by the oscilloscope. The position of the antenna was fixed while the TMS coil was positioned 60 mm above the antenna on the Z-axis (x = 0, y = 0, z = 60 mm) (Figure 1). The TMS coil irradiated single pulses (duration ~350 μs, repetition rate = 1 pulse/sec) at B field magnitude that varied from 10% to 100% of $B_{max}$ (1.2 T) using 10% incremental steps. Ten pulses were recorded for each B field magnitude and position; three independent experiments were completed on the Z-axis (n = 30). The antenna position was then moved along the X-axis during the TMS irradiation to gauge the lateral receiving signal. A similar voltage measurement was recorded when the antenna moved along the Y-axis. The distance of 60 mm between the TMS coil and the antennas was selected to guarantee stable electrophysiological measurements and allow sufficient manoeuvrability during the *in vivo* experiments. Copper loops were used in this part of the study as a proof of concept considering the moderate cytocompatibility of copper [30].

*Loop-antenna stimulation of nerves*

All work on animals followed protocols approved by the ethics committee of the School of Medicine at Western Sydney University. A total of 32 female Wistar rats weighing 287 ± 16 g were used for the *in vivo* experiments. The sciatic nerve was exposed and recording electrodes positioned following a standard approach (Supplementary 1.1). Prior to attempting the nerve activation using the TMS, the continuity of the sciatic nerve was tested by the production of CMAPs and a visible twitch of the muscle following electrical stimulation of the nerve with a constant current stimulator (pulse duration = 500 µsec, amplitude = 10 µA, 1 pulse per second). Once the continuity of the sciatic nerve was confirmed, the TMS coil was positioned 60 mm above the sciatic nerve with and without (control) the loop antenna placed around the nerve. The TMS delivered pulsed stimuli at a field magnitude of ~0.72 T (60% $B_{max}$). Each pulse was repeated 120 times with individual and averaged CMAP responses recorded using a computer program for further analysis (LabChart, Version 8.1.5). During some of these *in vivo* experiments the voltage induced in the antenna by the TMS was also measured (along with the action potential) to characterize the antenna performance. The same set-up described in the previous section was used to measure the voltage induced in the antenna surrounding the nerve. These measures were also repeated using a loop antenna that was insulated by a plastic coating to assess the importance of contact between the antenna copper wire and the nerve perineum for triggering action potentials. Similar measures, as described above, were completed in other rats; in this instance though, the compound action potential of nerves was measured instead of the muscle action potentials. The recording electrode was placed on the sciatic nerve 1 cm away from the loop antenna towards the foot, the reference electrode was fixed to the adjacent tissues, while the ground electrode was fixed to the leg skin of the rat. Measurements were restricted to TMS stimuli with an intensity of 0.72 T (60% of $B_{max}$). Animals were euthanized at the end of the procedure.

*Antenna currents*

The peak currents induced by the TMS in the loop antenna were measured using a custom-made ammeter (Supplementary 1.2). The loop antenna was placed around the sciatic nerves of Wistar rats and connected to the ammeter; currents were also measured when the nerve was not inside the loop antenna, maintaining the same set-up. Rats were kept under anaesthesia and all measures were recorded *in vivo*. Animals were euthanized at the end of the procedure.

*Compound Muscle Action Potential: AC versus DC currents*

The loop antenna was connected to the ammeter and placed around the sciatic nerve of Wistar rats. The TMS stimulated the nerves with the antenna and the CMAP was recorded. The loop antenna was then removed and replaced with a DC stimulator that delivered square pulses of 500 µsec duration with a current amplitude of 10 µA at a repetition rate of 1 second. The current intensity that triggered a CMAP similar to that elicited by the loop antenna was then recorded. Animals were euthanized at the end of the procedure.

*In vivo implantation of graft-antennas: stimulation of uncut nerves*

The graft-antenna was fabricated accordingly to a standard protocol that includes the formation of a chitosan adhesive film embedded with a gold strip (Supplementary 1.3). The graft-antenna was implanted on intact sciatic nerves of five rats (no nerve transection). After exposing the sciatic nerves (diameter ~1 mm), a graft-antenna measuring 5x5 mm was sterilised by dipping in 80% ethanol; it was then placed on a thin plastic backboard that was positioned underneath the sciatic nerve. The two edges of the adhesive were lifted from the backboard using microforceps and wrapped around the nerve such that the gold strip overlapped itself, forming a loop around and in contact with the nerve (Figure 4). The graft-antenna was bonded to the nerve by a laser emitting a power of 250 mW at 532 nm in a continuous wave, with a fibre core diameter of 200 nm [25–29]. The graft-antenna was spot-irradiated ensuring each spot (~6 mm) was irradiated for ~5 seconds; the spot was moved back and forth along the length of the graft-antenna, whilst gently rotating the nerve by manipulating the backboard, for a total of ~133 seconds. Care was taken not to irradiate the gold loop of the graft-antenna. Once the graft-antenna fully adhered to the nerve, forming a collar, the embedded gold strip became a loop antenna with diameter of ~1 mm. After implantation, the TMS coil stimulated the nerve via the gold loop at a distance of 6 cm from the graft-antenna, as described previously. The CMAP was recorded using the same set-up detailed before. After removing the plastic backboard, muscles and skin were closed using 3-0 Vicryl sutures and surgical staples, respectively. Topical antibacterial ointment and bittering agent were applied to the wound and rodents were returned to the animal facility in individual cages with no restriction of movement. The sciatic nerve of each rat was additionally stimulated once a week for 12 weeks by the TMS while the animal was under anaesthesia. The stimulation lasted 60 minutes and consisted of single pulses (duration ~350 μs, repetition rate = 1 pulse/s) at a field magnitude of ~0.72 T (60% $B_{max}$). Nerves were finally harvested before euthanizing animals, and histological samples were prepared (Supplementary 1.4). In separate experiments, the peak currents and voltages of the graft-antenna were also measured using the ammeter and oscilloscope set-ups described earlier.

*In vivo nerve grafting of transected nerves*

The rodents were prepared and sciatic nerves freed as described above. A 10 mm section of the sciatic nerve was then resected approximately 2-3 mm proximal to the trifurcation using microscissors. The nerve graft was removed, rotated in a distal-to-proximal and proximal-to-distal fashion, and replaced in its original position in the new orientation. The rats were randomly assigned to two groups.

*Graft-Antenna repair*

In this group, the proximal anastomosis was completed using the graft-antenna while the distal anastomosis was done with the adhesive alone (graft-antenna without the gold strip). The adhesive was laser-irradiated to activate tissue bonding and wounds were closed using the procedure described before. The graft-antenna was applied at the proximal anastomotic site in order to stimulate the intact part of the sciatic nerve, as illustrated in Figure 6.

*Adhesive repair*

Rats included in this group had the nerve graft reconnected using the adhesive alone at both the proximal and distal anastomotic sites. After surgery, all animals were placed in individual cages with no restriction of movement. To facilitate axon regeneration, the sciatic nerve of rats in the graft-antenna and adhesive-only group was stimulated by the TMS once a week for 8 weeks, as detailed before. Stimulation started a week after operation and at the end of the 8-week stimulation period, CMAP, CNAP and NCV were recorded to assess axon regeneration. Nerves were finally harvested before euthanizing animals and histological samples were prepared.

*Statistics*

All data are represented as mean ± SD; one-way ANOVA with Tukey's multiple comparison post-test and t-tests were used at a significance level of 0.05. Statistical analyses were performed using Prism 5 (GraphPad Software).

**Acknowledgements**

The Authors wish to thank Prof. André van Schaik and Prof. Kate Stevens for their project support.

**Author contributions**

A.L. conceived the loop stimulator and fabricated the graft-antenna. A.S. performed the electrophysiological experiments and antenna measurements with the help of D. Mahns, G.G., P.B., A.L. and G.T. The surgery was performed by Z.M. with the help of M.S., L.A., D. Mahns and A.S. The ammeter was built by G.G. Histology was done under the guidance of D.Mawad. Data were analysed by A.L., A.S., D. Mahns, P.B. and G.G. All authors contributed to and reviewed the submitted manuscript.

**Competing interests**

The authors declare there are no competing interests.

**Supplementary Information**

*Supplementary 1.1 (Surgery)*

Rodents were anaesthetized using 2% Isofluorane in 100% oxygen using a standard anaesthetic machine. Under sterile conditions, a 3-4 cm skin incision was made between the ischial tuberosity and the knee joint of the right leg. The plane between the gluteus maximus and biceps femoris muscles was identified and dissected using blunt dissection to expose ~1.5 cm of the sciatic nerve proximal its distal trifurcation. Under an Olympus operating microscope (1–40x), the nerve (diameter ~1 mm) was freed by dissecting surrounding connective tissue using a microscissor; care was taken to minimize nerve handling. The electrical response of the muscle was measured using a purpose-built AC coupled differential amplifier (100x gain, 1 hz High pass filter) and recorded using a digital to analog converter (Model 1401, Cambridge Electronic Design, UK). The recording electrode was placed in the rectus femoris muscle, the reference electrode was positioned to the adjacent tissues while the ground electrode was fixed to the leg skin of the rat. The relative position of electrodes was carefully measured with callipers for further data analysis.

*Supplementary 1.2 (Ammeter)*

A bespoke ammeter designed around the low noise instrumentation amplifier (INA118, Texas Instruments INC., USA) was used to measure the current flow induced in the loop antenna. The small current flow is transformed into a small voltage using a sampling resistor (nominal conversion factor: 1 µA = 1.2 mV). This resistor (1.2 Ω) is connected directly between the INA118 inputs and generates a voltage directly proportional to the current flow that is amplified by a factor of 1000 [V/V]. The amplified signal is directly acquired by the Powerlab system that it is used to record the action potential. The reference terminal of the INA118 is directly connected to the general grounding, which includes the rat and the full instrumentation, to avoid creation of ground loops. The bespoke ammeter is calibrated using a precise current source obtained with a 1/3 of REF200 (Texas Instruments INC., USA). Calibration can be verified by switching the current input of the sampling resistor to the calibrated constant current (100 µA) contained inside the REF200.

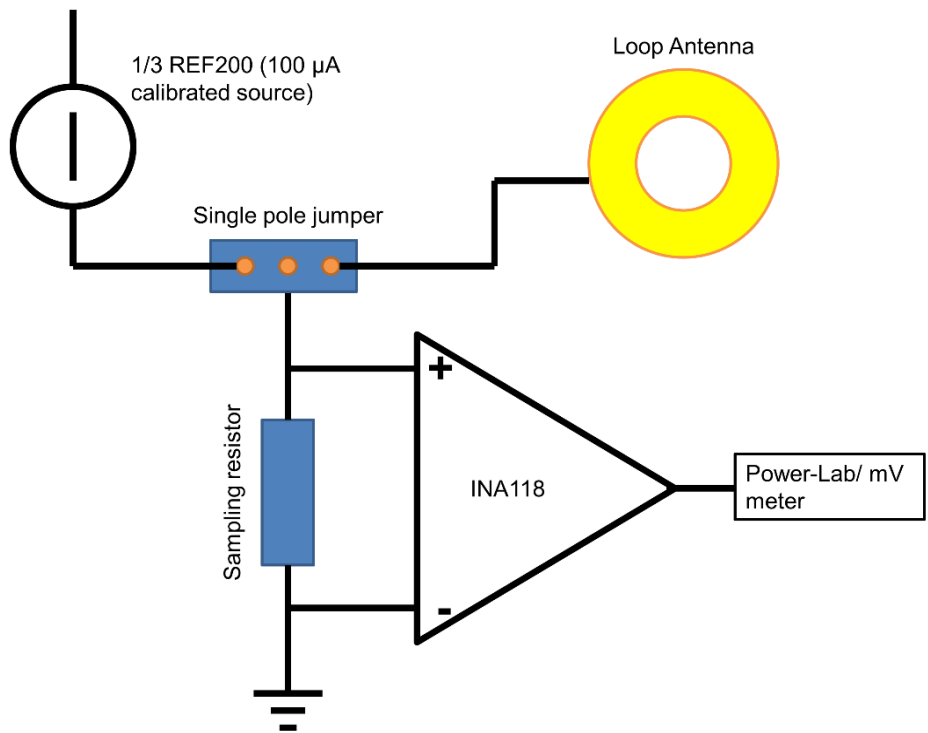

Simplified diagram of the bespoke ammeter (see text).

*Supplementary 1.3 (Graft-Antenna)*

The adhesive film was prepared using the method published earlier by our group [27,28]. Briefly, medium molecular weight chitosan (598 cps viscosity, 81% deacetylation; Sigma-Aldrich, Sydney, NSW, Australia) was dissolved at a concentration of 1.7% (w/v) in deionized water that contained 2% (v/v) acetic acid and 0.01% (w/v) rose bengal. The viscous solution was stirred for 14 days at room temperature (~25 °C) in the dark to avoid photo-bleaching of rose bengal. Insoluble matter was removed by centrifuging the rose bengal-chitosan solution at 3270×g for an hour. The collected supernatant was spread uniformly (~1.2 ml over ~12 cm$^2$) on a dry and sterile Perspex plate at room temperature. The solution was allowed to dry over 3 weeks which caused ~90% water content loss, forming a thin film which did not dissolve in water [29]. The rose bengal-chitosan film was carefully detached from the plate avoiding damage and small rectangular sections (~5x5 mm) were cut with scissors. An Emitech K550X gold coater (Quorum Emitech, East Sussex, England) sputtered a strip of gold onto the adhesive using a filter paper template. The chitosan adhesive was placed underneath the template and a gold strip was deposited with a width of 0.8 ± 0.1 mm and thickness of 50-80 nm. When this adhesive is placed around the nerve, the gold strip becomes a loop antenna that can receive electromagnetic radiation. The adhesive graft-antennas were stored in a sterile plastic box and kept in the dark at room temperature to avoid dye photobleaching.

*Supplementary 1.4 (Histology)*

Before euthanasia, sciatic nerves were exposed at the site of operation and inspected for neuroma formation, tissue adhesion and uncharacteristic inflammation. The nerves were then harvested in ~15 mm lengths and fixed in 5% paraformaldehyde solution in 0.1 M phosphate buffer for 24 hours at ~4 °C. Nerves were serially dehydrated in ethanol and embedded in paraffin. Transverse sections of 5–10 mm thickness were made 5 mm proximal and distal to the adhesive site. Samples were stained with Luxol Fast Blue to visualize myelinated axons and Haematoxylin and Eosin (H&E) to evaluate any adverse effect on nerves due to the TMS stimulation. Histological slides were scanned and analysed using an Aperio XT Slide Scanner (Aperio, Vista, CA, USA). Images were numbered and their identity concealed during analysis, which was carried out on ~55% of the cross-sectional area of the operated and non-operated (contralateral) nerves.

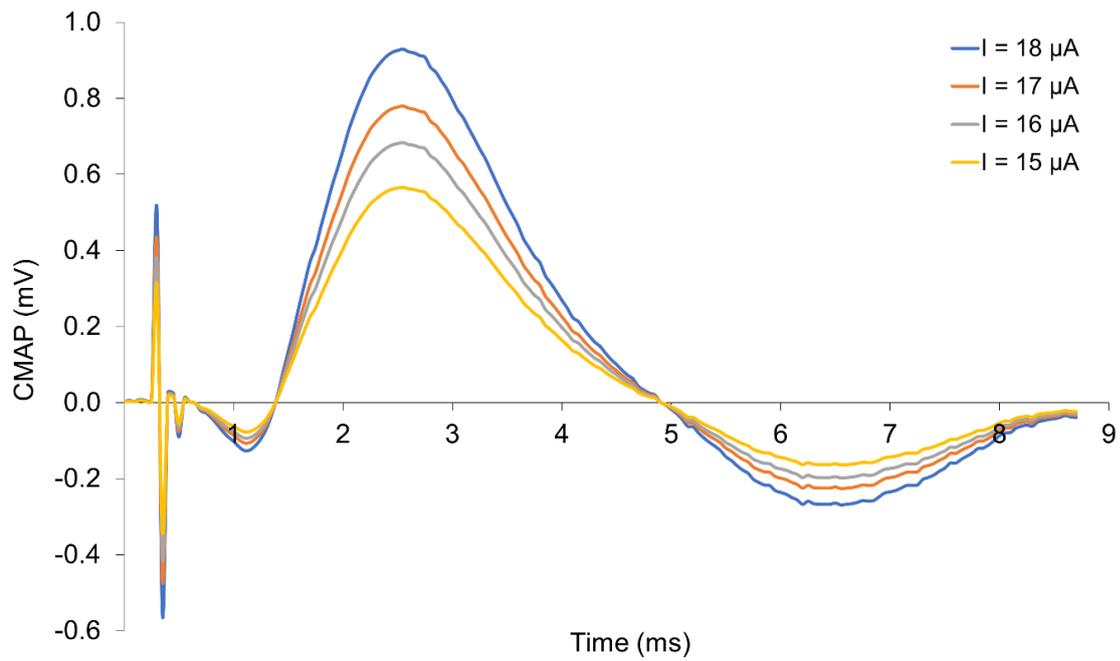

**Supplementary Figure 1.** Compound muscle action potentials (CMAPs) triggered in healthy sciatic nerves (controls) by a DC stimulator at different current levels. The signal amplitude decreases sharply from ~0.9 to ~0.5 mV when the current drops from 18 µA to 15 µA, respectively (n = 3).

**Supplementary Table 1. CMAP and Copper Loop Antenna Voltage**

|                       | Amplitude (mV)   | Loop Voltage (mV) |
|-----------------------|------------------|-------------------|
| CMAP + Oscilloscope   | 0.67 ± 0.09      | 10.49 ± 0.07      |
| CMAP                  | 0.70 ± 0.07      | -                 |
| CNAP + Oscilloscope   | 0.33 ± 0.06      | 10.49 ± 0.06      |
| CNAP                  | 0.33 ± 0.05      | -                 |

Legend. *CMAP + Oscilloscope*, Compound Muscle Action Potential amplitude measured when the loop antenna around the nerve is powered by the TMS. The loop antenna is connected to the oscilloscope. *CMAP*, CMAP amplitude measured when the loop antenna around the nerve is powered by the TMS without oscilloscope connection. *CNAP + oscilloscope*, amplitude of Compound Nerve Action Potential with antenna connected to oscilloscope. *CNAP*, amplitude of Compound Nerve Action Potential without oscilloscope connection. *Loop Voltage*, voltage induced in the copper loop antenna by the TMS. Three independent experiments were performed in each group (n = 3); 120 measures were averaged for the amplitude and voltage values in each experiment.

**Supplementary Table 2. Histomorphometric Results (Uncut Nerves)**

|                           | Proximal     | Distal       | Control      |
|---------------------------|--------------|--------------|--------------|
| **Myelinated Axon Count** | 1858 ± 75    | 1846 ± 81    | 1849 ± 71    |
| **Nerve Fiber Diameter (μm)** | 5.9 ± 1.6 | 5.8 ± 1.7 | 5.6 ± 1.5 |
| **Axon Diameter (μm)**    | 3.6 ± 1.5    | 3.6 ± 1.6    | 3.5 ± 1.5    |
| **Myelin Thickness (μm)** | 2.1 ± 0.8    | 2.1 ± 0.7    | 2.1 ± 0.6    |
| **Nerve Area (mm$^2$)**   | 54.3 ± 6.6   | 55.2 ± 9.2   | 54.7 ± 6.7   |

Histomorphometric results of the nerves 12 weeks after graft-antenna implantation. The TMS stimulated the nerves via the graft-antenna once a week for 1 hour (1 pulse/sec) at 60% $B_{max}$ (~0.72 T). The analysis was performed on ~55% of the total cross-sectional area of nerves (n = 5).

**Supplementary Table 3. Histomorphometric Results (Grafted Nerves)**

| Experimental Group            | Graft Antenna |           | Chitosan Adhesive |            | Control     |
|-------------------------------|---------------|-----------|-------------------|------------|-------------|
|                               | Proximal      | Distal    | Proximal          | Distal     |             |
| **Myelinated Axon Count**     | 1872 ± 100    | 1396 ± 68* | 1862 ± 75        | 1202 ± 66  | 1861 ± 79   |
| **Nerve Fiber Diameter (μm)** | 5.8 ± 1.8     | 5.4 ± 1.4 | 6.0 ± 1.7         | 5.1 ± 0.8  | 5.9 ± 1.8   |
| **Axon Diameter (μm)**        | 3.5 ± 1.6     | 3.2 ± 0.8 | 3.6 ± 1.5         | 3.1 ± 0.8  | 3.6 ± 1.7   |
| **Myelin Thickness (μm)**     | 2.2 ± 0.9     | 2.1 ± 0.8 | 2.3 ± 0.8         | 2.0 ± 0.7  | 2.2 ± 0.7   |
| **Nerve Area (mm$^2$)**       | 54.3 ± 8.9    | 54.2 ± 9.2 | 54.5 ± 6.0       | 55.6 ± 10.0 | 55.1 ± 7.9  |

Histomorphometric results of the nerves 8 weeks post-operatively; all nerves were briefly stimulated once a week for 1 hour (1 pulse/sec) and for 8 weeks by the TMS (~0.72 T). Nerves repaired with the graft antenna (n = 5) had more myelinated axons regrowing into the distal site than nerves operated with the adhesive only (one-way ANOVA, Tukey post-test, p = 0.0363). However, myelin thickness, fiber and axon diameter were not significantly different in the two groups.